\documentclass[aps,pra,reprint,superscriptaddress]{revtex4-1}

\usepackage{amssymb}		\usepackage{graphicx}		\usepackage{xspace}
\usepackage{color}
\usepackage{dcolumn}
\newcolumntype{.}{D{.}{.}{-1}}

\begin{document}

\let\oldSection\section

\renewcommand{\section}[1]{
\vspace{-11pt}
\oldSection{#1}
\vspace{-5pt}
}

\newcommand{\insertFigure}[4]{
\begin{figure}[t]
	\begin{center}
  	\includegraphics[width=#3]{#1}
  	\end{center}
  	\vspace{-1 em}
    \caption[#2]{#4}
\label{#2}
\end{figure}
}

\newcommand{\insertTwoImageFigure}[5]{
\begin{figure}[htbp]
	\begin{center}
  	\includegraphics[width=#4]{#1}
  	\includegraphics[width=#4]{#2}
  	\end{center}
    \caption[#3]{#5}
\label{#3}
\end{figure}
}

\newcommand{\figRef}[1]{Fig. \ref{#1}}

\newcommand*{\firstChange}[1]{#1}
\newcommand*{\secondChange}[1]{#1}
\newcommand*{\omittedChange}[1]{}

\newcommand{\snark}[1]{} 

\title{Impact of an introductory lab on students' understanding \\of measurement uncertainty}

\author{Benjamin Pollard}
\affiliation{Department of Physics, University of Colorado Boulder, Boulder, CO 80309, USA}

\author{Robert Hobbs}
\affiliation{Department of Physics, University of Colorado Boulder, Boulder, CO 80309, USA}
\affiliation{Department of Physics, Bellevue College, Bellevue, WA 98007, USA}

\author{Jacob T. Stanley}
\author{Dimitri R. Dounas-Frazer} 
\affiliation{Department of Physics, University of Colorado Boulder, Boulder, CO 80309, USA}
\author{H. J. Lewandowski}
\affiliation{Department of Physics, University of Colorado Boulder, Boulder, CO 80309, USA} 
\affiliation{JILA, National Institute of Standards and Technology and University of Colorado Boulder, Boulder, CO 80309, USA}

\begin{abstract}
Physics lab courses are an essential part of the physics undergraduate curriculum. Learning goals for these classes often include the ability to interpret measurements and uncertainties.  
The Physics Measurement Questionnaire (PMQ) is an established open-response survey that probes students' understanding of measurement uncertainty along three dimensions: data collection, data analysis, and data comparison. 
It classifies students'  reasoning into point-like and set-like paradigms, with the set-like paradigm more aligned with expert reasoning. 
In the context of a course transformation effort at the University of Colorado Boulder, we examine over 500 student responses to the PMQ both before and after instruction in the pre-transformed course.  
We describe changes in students' overall reasoning, measured by aggregating four probes of the PMQ. In particular, we observe large shifts towards set-like reasoning by the end of the course.
\end{abstract}

\pacs{}

\maketitle

\section{Introduction}
There is interest nationally in improving physics lab courses \cite{olson2012engage}, yet they remain under-studied compared to lecture classes \cite{national2012discipline}.
Proficiency with measurement and data analysis have been recently identified in national reports as important learning outcomes for physics lab courses \cite{kozminski2014aapt,Heron2016}.
An understanding of measurement uncertainty is critically tied to these learning outcomes.
Furthermore, measurement uncertainty has been identified by physics faculty at our own institution as an important learning goal in the context of an ongoing course transformation of the introductory lab class at the University of Colorado Boulder (CU).

Several assessments have been developed to study students' understanding of measurement uncertainty 
in physics lab classes, including the Concise Data Processing Assessment (CDPA), the Laboratory Data Analysis Instrument (LDAI), and the Physics Measurement Questionnaire (PMQ).
The CDPA was developed 
to measure student understanding of both measurement uncertainty and mathematical models of measured data \cite{Day2011}.
Subsequent work using the CDPA has focused on the use of scaffolding in instruction \cite{Holmes2014} and gender differences in physics labs \cite{Day2016}.
The LDAI was developed more recently to assess data analysis skills within the context of a single lab report, 
and highlights measurement uncertainty in that context \cite{Eshach2016}.
The CDPA and LDAI are both multiple choice assessments.
This work uses the Physics Measurement Questionnare (PMQ) \cite{Buffler2001, Volkwyn2008a}, which uses open-ended responses for greater insight into student reasoning.

The PMQ was developed by researchers at the University of Cape Town, South Africa and the University of York, United Kingdom.
Foundational work in Cape Town provided a theoretical basis for 
lab course design around measurement uncertainty \cite{Campbell2005}, highlighting the difference between conceptual understanding and procedural abilities, and finding that students can have one but not necessarily the other \cite{Buffler2001}. 
Later work showed that while traditional courses improved procedural performance, only a small fraction of students exhibited a deeper conceptual understading after instruction \cite{Volkwyn2008a}.
The PMQ has also been used at Uppsala University, Sweden, showing that a lab course about measurement 
yielded mixed results in students' ability to appropriately use ideas about uncertainty \cite{LippmannKung2006}.
In the United States, the PMQ has also been used
at the University of Maryland College Park \cite{Lippmann2003,Kung2005} and at North Carolina State University \cite{Abbott2003}, both showing significant gains in understanding of measurement uncertainty after completion of research-based courses.

The PMQ consists of several questions, or probes, concerning three aspects of understanding measurement uncertainty: data collection, data analysis, and data comparison.
Each probe comprises a decision about measurement uncertainty (usually posed as a multiple choice question) and an open-response text box for justifying or explaining that decision. 
Here we focus on four probes: \textit{repeating measurements} (RD), which probes data collection; \textit{using repeated measurements} (UR), which probes data analysis; and \textit{comparing same means with different spread} (SMDS) and \textit{comparing different means with the same spread} (DMSS), which both probe data comparison.
See Ref. \cite{Volkwyn2008a} for a more complete description of each probe.

The PMQ was designed to classify student reasoning into two broad paradigms: point-like and set-like.
Buffler et al. \cite{Buffler2009} defined these paradigms as follows.
Point-like reasoning stems from the idea that a single measurement can yield the ``true value" of a physical quantity.
It often results in individual measurements being considered independently of each other.
In contrast, set-like reasoning recognizes that no individual measurement yields the true value, and that multiple measurements will form a distribution.
The set-like paradigm is more aligned with expert-like reasoning.

In this work, we study students' understanding of measurement uncertainty in the context of an introductory lab physics course at CU.
We focus on two central research questions: (i) Can the PMQ provide significant signals that are sensitive to the range of student understanding in a large introductory lab physics course at our, or a similar, institution? and (ii) How well does our traditional physics introductory lab course teach measurement uncertainty concepts probed by the PMQ?
Our findings will inform ongoing course transformation work,
and form a baseline for studying the transformed course.

\section{Context and Methods}
This work focuses on the population of students that took a large-enrollment stand-alone introductory lab course at CU during Fall 2016.
The course is typically taken by students in their second semester of study, after completion of an introductory course on mechanics, and taken concurrently with an introductory course on electricity and magnetism.
Of all the students enrolled in the introductory lab course between 2004 and 2014, 76\% were male and 24\% were female; 74\% were white, 9\% were Asian-American, 8\% belonged to an underrepresented racial/ethnic group in physics in the US, and 4\% were international students \footnote{The CU Office of Planning, Budget, and Analysis, most recent available data.}.
In Fall 2016, 8\% of students in the course were physics majors, 63\% were engineering majors (excluding engineering physics), 18\% were non-physics science and math majors, and 11\% were majors in other disciplines.

The course is structured as a series of six two-week lab activities.
Activities cover a range of topics in mechanics, electricity and magnetism, and other areas of physics.
Students write lab reports after completing each activity.
The course also includes a 50 minute weekly lecture with clicker questions, homework assignments on error analysis, and pre-lab questions. 
It does not have a midterm or a final exam.

In this study, the PMQ was administered electronically 
at the beginning and the end
of an introductory physics lab course described above.
Both the pre-test and the post-test acted as in-class assignments in the course, graded only for participation and constituting a combined ~2\% of the final course grade.
Of the 588 students who completed the course, 525 completed both the pre- and post-tests and formed the data set for our analysis.

Student open-ended responses were initially coded by one of the authors (R.H.) using an expanded version of the codebook developed by Volkwyn et al. \cite{Volkwyn2005}.
These codebooks included a separate set of codes for each probe.
Additional emergent codes were added to Volkwyn's set during the initial coding process.
After the responses were coded, the research team collaboratively assigned a paradigm designation to each code that was indicative of the type of reasoning used: either \textit{P} for point-like, \textit{S} for set-like, or \textit{N} for neither.
The \textit{N} designation encompassed reasoning that did not fit into the point-set paradigms, as well as responses that were off-topic, uninterpretable, or too vague to determine the underlying reasoning of the student.
A single response could be assigned multiple codes, however almost all of the responses fell within a single paradigm.
The few responses (2.5\% of responses to all pre- and post-test probes) with multiple codes that mapped to multiple paradigms were counted as \textit{N}.

To determine the reliability of our paradigm assignments, a test of inter-rater reliability was performed for each probe. 
A random subset of 10\% of the pre-test responses was coded independently by another one of the authors (B.P.), and the associated paradigms for each response were then compared to the paradigms from the first rater.
Two measures of inter-rater agreement were computed.
The percent agreement was 78\%, and the Cohen's kappa statistic was 0.63, indicating ``substantial agreement" \cite{Cohen1960,Blackman2000}.

To measure overall student understanding of measurement uncertainty, we count the number of \textit{P} or \textit{S} responses a student gave to each of the four probes used from the PMQ.
Additionally, in order to simplify interpretation of the overall trends in our data set, we 
developed a well-defined method of combining the paradigm designations from the individual PMQ probes into a single paradigm designation for a student overall.
Assigning overall student paradigms has also been done in previous studies using the PMQ at Cape Town and at Maryland
\cite{Lippmann2003, Volkwyn2008a}; studies at Uppsala \cite{LippmannKung2006} and North Carolina \cite{Abbott2003} did not assign paradigms to students overall.
Here, we refer to the overall student paradigms as \textit{point-like} and \textit{set-like}, using italics to differentiate these coding designations from the more general and conceptual idea of point-like or set-like reasoning. 

\begin{table}[]
\centering
\caption{Definitions of overall student paradigms.}
\label{pdgmDefTable}
\begin{ruledtabular}
\begin{tabular}{l . .}
\multicolumn{1}{l}{Student paradigm}     & \multicolumn{1}{c}{Number of \textit{P}'s}  & \multicolumn{1}{c}{Number of \textit{S}'s}   \\ \hline \noalign{\smallskip}
\textit{point-like} & \geq1						 & 0       						\\ 
\textit{set-like}   & 0       						 & \geq1 						\\  
\textit{mixed}    & \geq1 						 & \geq1 						\\  
\textit{mixed}    & 0        					 & 0       						\\  
\end{tabular}
\end{ruledtabular}
\end{table}
Our method was as follows, and is also summarized in Table \ref{pdgmDefTable}.
If a student's responses to the four probes from the PMQ (for either the pre-test or the post-test) included \textit{P} but no \textit{S} designations, we classified that student as \textit{point-like} (for the pre-test or the post-test).
Within one standard deviation, our average \textit{point-like} student had about 1--3 \textit{P} responses.
If a student's responses included \textit{S} but no \textit{P} designations, we classified that student as \textit{set-like}.
Our average \textit{set-like} student had about 2--4 \textit{S} responses.
If a student's responses included both \textit{P} and \textit{S} designations, we classified that student as \textit{mixed}.
And finally, if a student's responses were exclusively designated as \textit{N}, with no \textit{P} or \textit{S} responses, we classified that student as \textit{mixed} as well.
However, that final case occurred only twice in the entire data set, once in the pre-test and once in the post-test, albeit from different students.
Our average \textit{mixed} student had about 1--2 \textit{P} responses and about 1--3 \textit{S} responses.

We used two measures of significance when comparing data in this study to investigate changes between the pre-test and post-test distributions of student paradigms: the nonparametric Mann-Whitney U-test at the 5\% significance level as a test of statistical significance \cite{Mann1947}, and the variance of the multinomial distribution as an indicator of practical significance.
We used Mann-Whitney on distributions produced by three different ways of measuring student understanding: by counting the number of \textit{P} responses for each student, by counting the number of \textit{S} responses for each student, or by treating each overall student paradigm (\textit{point-like}, \textit{set-like}, or \textit{mixed}) as a distinct entity. 
We used the variance of the multinomial distribution to calculate uncertainties in the fraction of students with a given paradigm.
We used the variance, normalized to the number of students in the population, to calculate the 95\% confidence interval.

\insertFigure{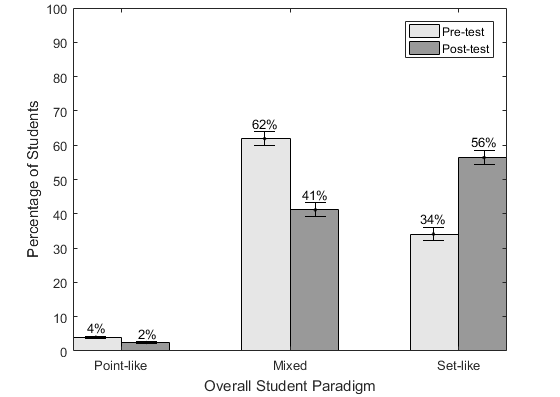}{ShiftsByStudent}{\linewidth}{
Overall understanding of measurement uncertainty by overall student paradigm for both the pre-test and the post-test.
Error bars show 95\% confidence interval.
By treating the overall student paradigms together as a single distribution, the pre-test and post-test distributions are statistically different ($p \ll 0.01$).
}
\section{Results and Discussion}
Figure \ref{ShiftsByStudent} shows the distribution of overall student paradigms both before and after instruction. 
We see a significant change in paradigm distribution after instruction ($p \ll 0.01$), with an increase in the percentage of students coded as \textit{set-like} and a decrease in the percentage of those coded as \textit{point-like} and \textit{mixed}.
Furthermore, there were very few overall \textit{point-like} students in both the pre-test and the post-test.
This low incidence of consistent point-like reasoning is similar to results from Maryland \cite{Lippmann2003}, but contrasts with previous studies using the PMQ in Cape Town \cite{Buffler2001,Campbell2005}.
The student populations and institutional contexts in Cape Town differ in many ways from Colorado and Maryland, the latter two of which are predominantly white, large, selective R1 institutions with large physics programs in the United States.
Nonetheless, we measure a statistically and practically significant decrease in the percentage of students coded as \textit{point-like} after instruction at CU. 
Moreover, the prevalence of \textit{mixed} students
suggests that most students come in with an inconsistent and context-dependent understanding of measurement uncertainty, and that such mixed understandings remain for a considerable number of students after instruction.

\insertFigure{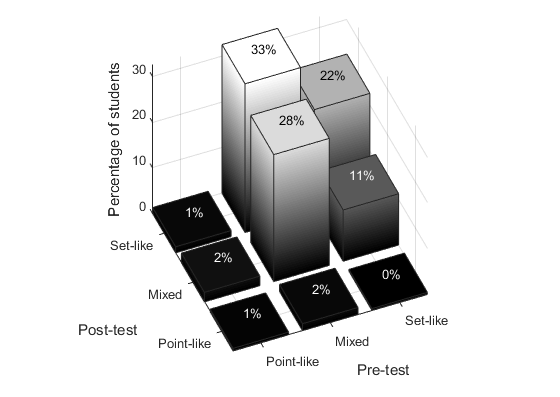}{PrePost3D}{\linewidth}{
Shifts in overall student understanding of measurement uncertainty between pre-test and post-test by overall student paradigm. 
}
To further explore how students' understanding shifted during instruction, we show a two-dimensional histogram of overall student paradigms in Fig. \ref{PrePost3D}.
Each bar represents the subpopulation of students that started in a specific paradigm for the pre-test, and then ended in a specific paradigm for the post-test.
Since our population was dominated by students showing \textit{set-like} or \textit{mixed} reasoning, we focus on those paradigms here.
To better understand the changes in understanding that led to these shifting distributions in overall paradigm, we look for significant changes between pre-test and post-test. 
We also calculate the average change in the number of \textit{P} and \textit{S} responses for each prominent subpopulation in Fig. \ref{PrePost3D}.

The populations of students that did not change their overall paradigm, either remaining \textit{set-like} or remaining \textit{mixed} after instruction, did not have significantly different numbers of individual \textit{S} responses between pre-test and post-test ($p = 0.31$ for \textit{set-like} and $p = 0.09$ for \textit{mixed}).
However, the populations of students who did change their overall paradigm, from \textit{set-like} to \textit{mixed} or \textit{mixed} to \textit{set-like}, had significantly different numbers of individual \textit{S} responses between pre-test and post-test ($p \ll 0.01$ in both cases).
The students who switched from \textit{mixed} to \textit{set-like} had an average shift in the number of individual \textit{S} responses of 0.97 $\pm$ 0.09 and an average shift in the number of \textit{P} responses of -1.09 $\pm$ 0.05, gaining one \textit{S} response and losing one \textit{P} response on average.
Similarly, the students who switched from \textit{set-like} to \textit{mixed} had an average shift in the number of individual \textit{S} responses of -0.82 $\pm$ 0.15 and an average shift in the number of \textit{P} responses of 1.09 $\pm$ 0.09, losing one \textit{S} response and gaining one \textit{P} response on average.

Lastly, we note a limitation to our analysis at the level of overall student paradigms. 
While this approach gives a comprehensive picture of general student understanding, 
the 
prevalence of \textit{mixed} reasoning suggests that there are marked differences in students' understanding between individual probes.
A deeper look 
probe-by-probe, 
both comparing probes to each other and looking at pre-post shifts on a probe-by-probe basis, would yield insight not captured by the analysis presented here.

\section{Conclusions and Outlook}
This work answers two overall research questions, as discussed above.
With regard to question (i), we find that the PMQ produces significant signals in the context of a large introductory lab physics course at CU.
It captures 
a variety of overall paradigms of students' understanding of measurement uncertainty.
Furthermore, it shows statistically significant shifts towards more expert-like reasoning between pre-test and post-test distributions.
Additionally, there are notable differences between results from our student population and those measured in previous PMQ studies, in particular a very low incidence of consistently point-like reasoning both before and after instruction.
Despite the low number of consistent point-like reasoners, the PMQ nevertheless provides useful information for a large standalone lab course at an R1 university in the US. 
Student populations with different levels of preparation compared to CU students may 
include a different fraction of consistent \textit{point-like} reasoners prior to instruction.
Nonetheless, developing new probes, or modifying existing ones, could improve signals from the PMQ when used at CU or similar institutions.

With regard to question (ii), using the PMQ, we measure significant shifts towards set-like reasoning after instruction in our traditional physics introductory lab course.
However, the majority of our students initially show an inconsistent understanding of set-like concepts across individual probes of the PMQ, and a sizable fraction of them continue to show such inconsistency after instruction.
These results establish a baseline that we will use to inform and evaluate 
our introductory lab course transformation.
Future investigation 
will verify whether these trends hold 
in additional semesters.
Furthermore, looking in detail at individual PMQ probes will help us to understand which aspects of uncertainty yield 
shifts, either positive or negative, after instruction.
This information could aid in designing instructional approaches that leverage the existing set-like concepts that students bring into our class, and move students towards more consistent expert reasoning.
In this way, our subsequent course transformation will perhaps yield greater shifts towards set-like reasoning, and more consistent reasoning across different aspects of measurement uncertainty. \\

\section{Acknowledgments}
We acknowledge Bethany Wilcox for invaluable discussions
on quantitative analysis. 
This material is based upon work supported by the NSF under Grant No. PHY-1125844, and by the University of Colorado.

\bibliography{references}

\end{document}